\documentclass[twocolumn,
showpacs,preprintnumbers,amsmath,amssymb,floatfix,aps]{revtex4}

\usepackage{graphicx}
\usepackage{dcolumn}
\usepackage{bm}

\begin{document}


\title{ Molecular signatures in the structure factor of an interacting Fermi gas.} 

\author{R. Combescot$^{a}$, S. Giorgini$^{b}$ and S. Stringari$^{b}$}
\affiliation{$^{a}$Laboratoire de Physique Statistique, Ecole Normale Sup\'erieure$^\star$,  
24 rue Lhomond, 75231 Paris Cedex 05, France \\
$^{b}$Dipartimento di Fisica, Universit\`a di Trento and CNR-INFM BEC Center, I-38050 Povo, Trento, Italy}

\begin{abstract}
The static and dynamic structure factors of an interacting Fermi gas along the BCS-BEC crossover are calculated at 
momentum transfer $\hbar{\bf k}$ higher than the Fermi momentum. The spin structure factor is found to be very sensitive 
to the correlations associated with the formation of molecules. On the BEC side of the crossover, even close to 
unitarity, clear evidence is found for a molecular excitation at $\hbar^2 k^2 /4m$, where $m$ is the atomic mass. 
Both quantum Monte Carlo and dynamic mean-field results are presented.
\end{abstract}
\pacs{Valid PACS appear here}

\maketitle

The possibility of producing weakly bound molecules in interacting ultracold Fermi gases, raises very interesting 
challenges. These molecules are formed near a Feshbach resonance for positive values of the s-wave scattering length, 
have bosonic character and exhibit Bose-Einstein condensation at low temperature~\cite{Exp,Bourdel}. They have a remarkably 
long life-time as a consequence of the fermionic nature of the constituents which quenches the decay rate associated with 
three-body recombinations. Several properties of this new state of matter have been already investigated experimentally
in harmonically trapped configurations. These include the molecular binding energy~\cite{Regal}, the release 
energy~\cite{Bourdel}, the size of the molecular cloud~\cite{Bartenstein}, the frequency of the collective 
oscillations~\cite{Kinast}, the pairing energy~\cite{Chin}, the vortical configurations~\cite{Zwierlein} and the 
thermodynamic behavior~\cite{Thomas}.

In this paper we investigate another feature of these new many-body configurations, directly related to the molecular 
nature of the constituents: the behavior of the static and dynamic  structure factor at relatively high momentum transfer. 
Experimentally the structure factor can be measured with two-photon Bragg scattering where two slightly detuned laser beams 
are impinged upon the trapped gas. The difference in the wave vectors of the beams defines the momentum transfer 
$\hbar {\bf k}$, while the frequency difference defines the energy transfer $\hbar \omega$. The atoms exposed to these beams 
can undergo a stimulated light scattering event by absorbing a photon from one of the beams and emitting a photon into the 
other. This technique, which has been already successfully applied to Bose-Einstein condensates~\cite{Bragg}, provides direct 
access to the imaginary part of the dynamic response function and hence, via the fluctuation-dissipation theorem, to the 
dynamic structure factor. At high momentum transfer the response is characterized by a quasi-elastic peak at 
$\omega=\hbar k^2/2M$, where $M$ 
is the mass of the elementary constituents of the system. The position of the peak is consequently expected to depend on 
whether photons scatter from free atoms ($M=m$) or  molecules ($M=2m$). For positive values of the scattering length both 
scenarios are possible and their occurrence depends on the actual value of the momentum transfer. If $k$ is much larger than 
the inverse of the molecular size, photons mainly scatter from  atoms and the quasi-elastic peak takes place at $\hbar k^2/2m$. 
In the opposite case, photons scatter from molecules and the excitation strength is concentrated at $\hbar k^2/4m$. The two 
regimes are associated with different velocities of the scattered particles given, respectively, by $\hbar k/m$ and $\hbar k/2m$. 

Let us suppose that the Fermi gas consists of an  equal number $N/2$ of atoms in two hyperfine states 
(hereafter called spin-up and spin-down). Using 
$S_{\uparrow\uparrow}(k,\omega )=S_{\downarrow\downarrow}(k,\omega )$ and 
$S_{\uparrow\downarrow}(k,\omega )=S_{\downarrow\uparrow}(k,\omega )$, we can write the $T=0$ dynamic structure factor in the 
form  
\begin{equation} \label{S}
S(k,\omega)= 2\left(S_{\uparrow\uparrow}(k,\omega) + S_{\uparrow\downarrow}(k,\omega) \right)\, , 
\end{equation}
with 
\begin{equation} \label{Sdown}
S_{\sigma\sigma^\prime}(k,\omega)\! =\! \sum_n \!<0|\rho_\sigma(k)|n><n|\rho_{\sigma^\prime}^\dagger(k)|0> 
\delta(\hbar \omega - E_{n0})
\end{equation} 
where $\rho_{\sigma}(k)=\sum_{i_\sigma} e^{-ikz_i}$  are the spin-up ($\sigma=\uparrow$) and spin-down ($\sigma=\downarrow$) 
components of the Fourier transform of the {\it atomic} density operator, while $|n>$ and $E_{n0}=E_n-E_0$ are the eigenstates 
and eigenenergies of the many-body Hamiltonian $H$. 

The frequency integral of the dynamic structure factor defines the so-called static structure factor 
relative to the different spin components:
\begin{equation}
\label{Sstatic}
\hbar\int_0^{\infty} d\omega  \, S_{\sigma\sigma^\prime}(k,\omega)= \frac{N}{2} S_{\sigma\sigma^{\prime}}(k)\, .
\end{equation}
Using the completeness relation one can write
\begin{equation}
\label{Sstatic2}
S_{\sigma\sigma^{\prime}}(k) = \frac{2}{N} <0|\sum_{i_{\sigma},j_{\sigma^{\prime}}}e^{-i(k(z_i-z_j)}|0>\, .
\end{equation}
The total spin structure is then given by  
$S(k) = N^{-1}\hbar \int d\omega \,S(k,\omega)= S_{\uparrow\uparrow}(k)+ S_{\uparrow\downarrow}(k)$. The 
static structure factor is related to the two-body correlation functions through the relationships 
\begin{eqnarray}
S_{\uparrow\uparrow}(k) &=& 1+\frac{n}{2}\int d{\bf r}\,[g_{\uparrow\uparrow}(r)-1]e^{i{\bf k \cdot r}}
\nonumber\\
S_{\uparrow\downarrow}(k) &=& \frac{n}{2}\int d{\bf r}\,[g_{\uparrow\downarrow}(r)-1]e^{i{\bf k \cdot r}}\;,
\label{corr}
\end{eqnarray}
yielding $S(k) = 1+n\int d{\bf r}\,[g(r)-1]e^{-i{\bf k \cdot r}}$ with 
$g(r)=[g_{\uparrow\uparrow}(r)+g_{\uparrow\downarrow}(r)]/2$. In the above equations $n$ is the total particle density fixing the
Fermi wave vector according to $k_F^3=3\pi^2n$. The behavior of the structure factor $S(k)$ at 
small momenta ($k\ll k_F$) is dominated by long-range correlations which give rise to a linear dependence in $k$. In a 
superfluid the slope is fixed by the sound velocity $c$, through the general law $S(k) = k/2mc$. In this paper we are 
however mainly interested in the behavior at large momentum transfer, typically such that $k \gtrsim k_F$. 

In the limit of very large $k$, the sum in Eq.(\ref{Sstatic2}) is dominated by the autocorrelation term $i=j$ with identical 
spins. This leads to $S_{\uparrow\uparrow}(k) \rightarrow 1$. On the other hand, 
since there is no autocorrelation with different spins, $S_{\uparrow\downarrow}(k) \rightarrow 0$ for very large $k$. Actually 
in the ideal Fermi gas the dynamic spin-up spin-down structure factor identically vanishes ($S_{\uparrow\downarrow}(k,\omega)=0$) 
for all values of $k$ and $\omega$, reflecting the complete absence of correlations between particles of  opposite 
spin~\cite{Pines}. This quantity is therefore particularly well suited for studying the effect of interactions. 
Let us first consider the case of small and positive scattering length $k_F a\ll 1$. This is the so-called Bose-Einstein 
condensation (BEC) regime, where we have a dilute gas of weakly bound molecules, made of atoms with opposite spins, with 
normalized wavefunction $\Phi_0({\bf r})$ for the relative motion. When we consider distances of the order of the molecule 
size $a$, we have naturally a strong correlation between opposite spin atoms belonging to the same molecule. In this case the 
sum (\ref{Sstatic2}) is dominated by this contribution, which gives:
\begin{equation} 
\label{SupdownM}
S_{\uparrow\downarrow}(k) = \int d{\bf r} \;e^{i{\bf k \cdot r}}\,n_{mol}({\bf r})\;, 
\end{equation} 
where $n_{mol}({\bf r})=|\Phi_0({\bf r})|^2$ is the probability to have the atoms separated by ${\bf r}$. This holds only for 
$k \gg k_F$, otherwise one has to take into account also correlations between atoms belonging to different molecules. In particular 
one finds that, for $ k_F \ll k \ll 1/a$, $S_{\uparrow\downarrow}(k)=1$, so that $S(k)=2$. This corresponds to the regime where the 
elementary constituents ``seen" by the scattering probe are molecules and not atoms. 

\begin{figure}
\begin{center}
\includegraphics*[width=7.5cm]{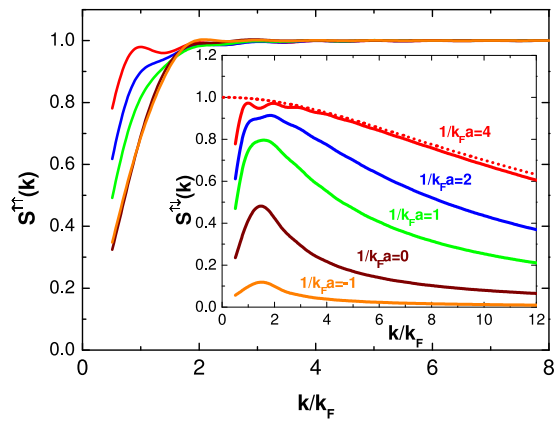}
\caption{(color online). Spin-dependent static structure factor $S_{\uparrow\uparrow}(k)$ and $S_{\uparrow\downarrow}(k)$ 
(inset) obtained using the QMC method, for different values of the interaction strength as indicated in the figure. The red 
dotted line in the inset corresponds to the Fourier transform of $n_{mol}(r)$ [Eq. (\ref{SupdownM})].}
\label{fig1}
\end{center}
\end{figure}

When one moves away from this BEC regime towards the resonance, where $k_Fa \gg 1$, the many-body wave function cannot be simply 
written in terms of  molecules anymore. In this very interesting regime the function $S_{\uparrow\downarrow}(k)$ is smaller than 
unity, but can still significantly differ from zero, reflecting the crucial role played by interactions (see Fig.~\ref{fig1}).

\begin{figure}
\vspace{-6mm}
\begin{center}
\includegraphics*[width=7.5cm]{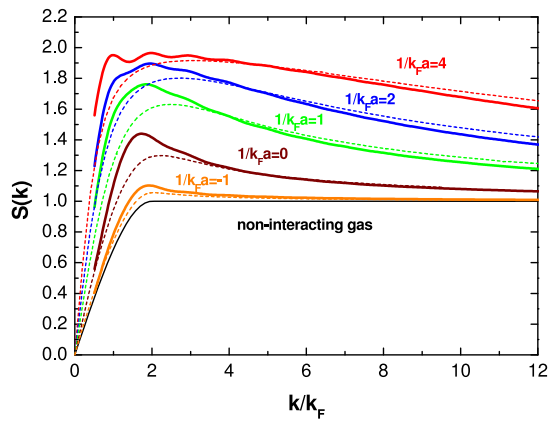}
\caption{(color online). Total static structure factor $S(k)$ for different values of the interaction strength. Solid lines 
correspond to QMC results and dashed lines to BCS results. The black line refers to the non-interacting gas.}
\label{fig2}
\end{center}
\vspace{-10mm}
\end{figure}

\begin{figure}
\begin{center}
\includegraphics*[width=7.5cm]{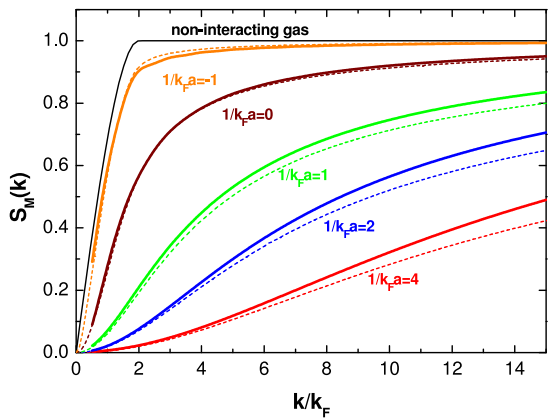}
\caption{(color online). Magnetic static structure factor $S_M(k)$ for different values of the interaction strength. 
Solid lines correspond to QMC results and dashed lines to BCS results. The black line refers to the non-interacting gas.}
\label{fig3}
\end{center}
\vspace{-10mm}
\end{figure}

In Figs.~\ref{fig1}-\ref{fig3} we report quantum Monte Carlo (QMC) results of the static structure 
factors as a function of $k/k_F$, calculated for different values of the relevant interaction parameter $k_Fa$. The simulations 
are carried out following the method described in Ref.~\cite{MC}. The direct output of the calculation are the spin-dependent 
pair correlation functions, $g_{\uparrow\uparrow}(r)$ and $g_{\uparrow\downarrow}(r)$. The corresponding structure factors are 
obtained through the Fourier transformation of Eq. (\ref{corr}). Notice that the low-momentum behavior of the structure factors 
can not be accessed in the QMC simulation due to the finite size of the simulation box. For our system ($N$=66 particles 
and periodic boundary conditions) a reliable calculation of $S_{\sigma\sigma^\prime}(k)$ is limited to $k\ge 0.5\, k_F$.

The results for  $S_{\uparrow\downarrow}(k)$ for small positive values of $k_Fa$ (see Fig.~\ref{fig1}) are  fully consistent with 
the transition from the molecular regime, where $S_{\uparrow\downarrow}(k)\simeq 1$, to the atomic one where 
$S_{\uparrow\downarrow}(k) \to 0$ at large $k$. The transition is well described by Eq. (\ref{SupdownM}) which is reported in 
the inset of Fig.~\ref{fig1} for the value $k_Fa=1/4$. The molecular density profile $n_{mol}({\bf r})$ entering 
Eq. (\ref{SupdownM}) has been calculated using the same two-body potential of range $R_0$ employed in the QMC 
simulation~\cite{notemolecule}. 
In the large-$k$ region, $1/R_0\gg k\gg \max(k_F,1/a)$, the decay of $S_{\uparrow\downarrow}(k)$ is proportional to $1/k$ as a 
consequence of the $1/r^2$ short-range behavior of $g_{\uparrow\downarrow}(r)$. This behavior is fixed by two-body physics 
which dominates at short distances. The proportionality coefficient of the $1/k$ law is given by 
$2 \pi ^{2} \,{\rm lim}_{r \rightarrow 0} \,[r^{2} |\Phi_0({\bf r})|^2]$ in the deep BEC regime ($r\to 0$ here means 
$R_0\ll r\ll a$) and by many-body effects closer to the resonance and in the BCS regime.
For example, at unitarity ($1/k_Fa=0$), we find $S_{\uparrow\downarrow}(k)\sim 0.85 k_F/k$. 
For $1/k_Fa\gg 1$, the ``atomic" regime, where  $S_{\uparrow\downarrow}(k)$ approaches zero, is reached only at 
extremely large values of $k/k_F$ reflecting the crucial role played by interactions on this quantity even at high $k$.
Conversely $S_{\uparrow\uparrow}(k)$ is less sensitive to interactions unless one considers small values of $k$. 
This quantity approaches quite rapidly its large-$k$ limit $S_{\uparrow\uparrow}(k) \to 1$ (see Fig.~\ref{fig1}). 

The results 
for the total structure 
factor are shown in Fig.~\ref{fig2}. In the BEC regime $S(k)$ first increases and reaches a plateau where $S(k)\simeq 2$, 
which is characteristic of the molecular regime. At higher $k$ it decreases and eventually approaches the uncorrelated 
value $S(k)=1$. In Fig.~\ref{fig3} we show the results for the magnetic structure factor 
$S_M(k)= S_{\uparrow\uparrow}(k) - S_{\uparrow\downarrow}(k)$. Compared to the total structure factor this quantity in the 
BEC regime is strongly quenched in the range $ka \ll1$ where it behaves like $(ka)^2$. In fact, the leading contributions 
of $S_{\uparrow\uparrow}(k)$ and $S_{\uparrow\downarrow}(k)$, of order of $1$, cancel each other in this regime. 

In Figs.~\ref{fig2}-\ref{fig3}  we have also reported the results of the dynamic mean-field theory 
(see discussion below) which 
provides very reasonable predictions for the structure factors as appears from the comparison with the microscopic findings of  
the QMC simulation. The deviations at large $k$ in the BEC regime are mainly due to the details of the molecular wave function 
used in the QMC simulation at distances $r\lesssim R_0$.

\begin{figure}
\begin{center}
\includegraphics*[width=8.5cm]{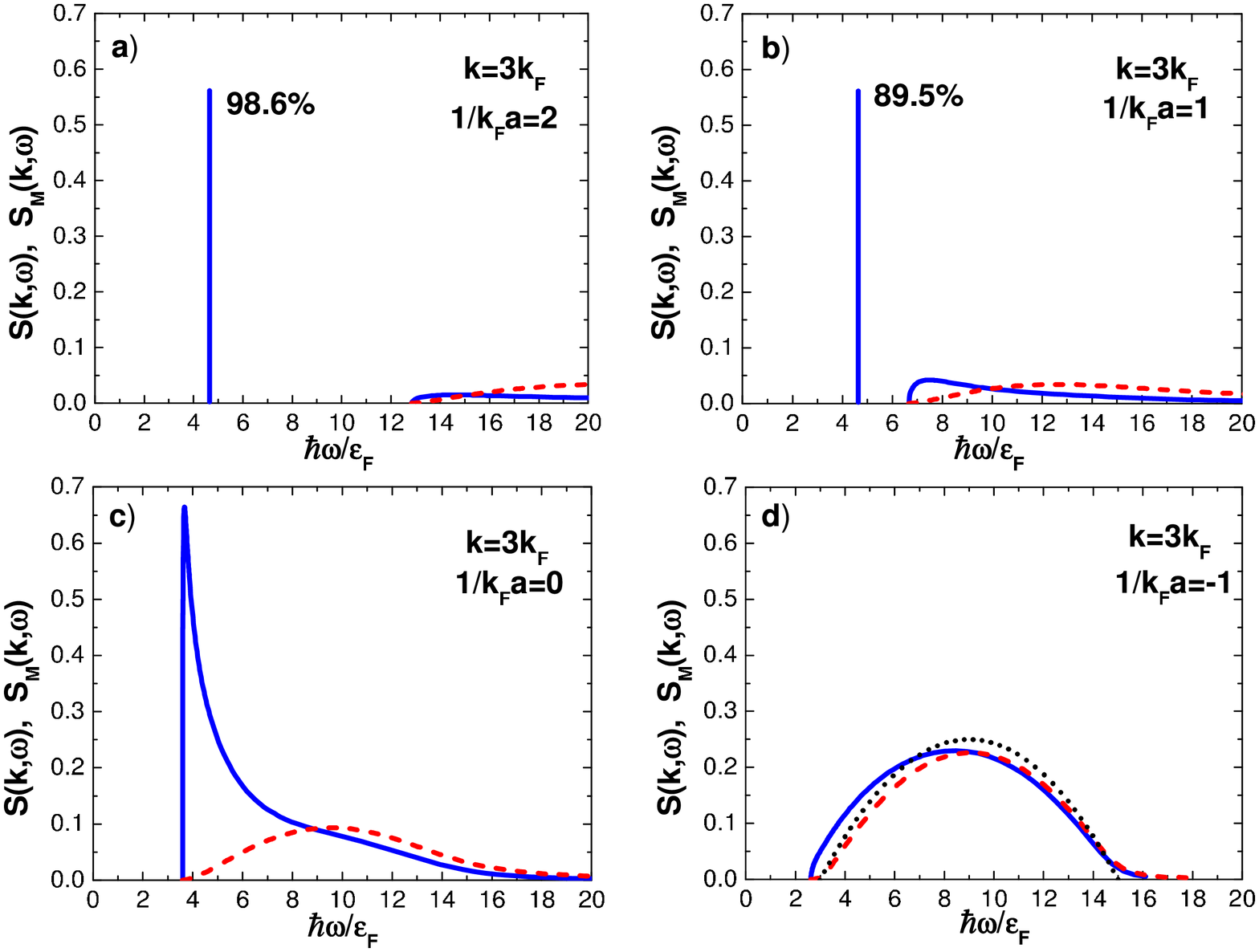}
\caption{(color online). Dynamic structure factor, from the BCS approach, as a function of $\omega$ for fixed 
momentum transfer $k=3k_F$ and for various interaction strengths. Blue solid lines refer to the total structure
factor $S(k,\omega)$ and red dashed lines to the magnetic one $S_M(k,\omega)$. The black dotted line in panel d) corresponds 
to the non-interacting gas. In panel a) and b) is also indicated the weight of the sharp peak to $S(k)$. The units of 
the dynamic structure factor are $N/(2\epsilon_F)$.}
\label{fig4}
\end{center}
\vspace{-10mm}
\end{figure}

Let us now discuss the behavior of the dynamic structure factor. Since QMC calculations for this quantity are not available, 
we will discuss here the predictions of the dynamic self-consistent mean-field approach based on the Bardeen-Cooper-Schrieffer 
(BCS) theory, which has proven to be quite accurate in the evaluation of the static structure factors (see Figs.~\ref{fig2}, 
\ref{fig3}). To evaluate the dynamic structure factor in the self-consistent BCS theory a 
convenient procedure is given by the calculation of the linear response function. This is obtained by writing a kinetic 
equation which generalizes the well known Landau equation for Fermi liquids to the superfluid state. This equation includes
as driving terms the off-diagonal self-consistent field, which ensures that the resulting theory is gauge invariant, in contrast 
to the original elementary BCS theory. In particular this approach leads naturally to the appearance of the gapless 
Bogoliubov-Anderson mode on the BCS side of the crossover~\cite{Anderson}, which goes continuously into the Bogoliubov  mode in 
the opposite BEC regime~\cite{BCS-BEC}. On the BCS side this collective mode exists only at low wavevector $k$.  
When $k$ is increased it merges at some 
point into the continuum of single-particle excitations. By approaching the Feshbach resonance from the BCS side, the threshold 
of single-particle excitations is pushed towards higher energies, which makes the merging occur for higher $k$. Rather soon, on 
the other side of the resonance, i.e. for small positive $1/k_Fa$, the mode dispersion relation never meets the continuum 
anymore, and it evolves to the standard Bogoliubov mode of the molecular BEC regime, the continuum of the single-particle 
excitations being located at higher energies~\cite{Roland}. At the high momenta considered in this work the Bogoliubov mode takes 
the free 
molecule dispersion  $\omega(k)= \hbar k^2/4m$. This result  is fully consistent with the behavior of the static structure 
factor $S(k)$ discussed above. In fact using the f-sum rule 
\begin{equation} \label{fupup}
\hbar^2\int \!d\omega \;\omega S(k,\omega)= N\;\frac{\hbar^2 k^2}{2m} \;, 
\end{equation}  
and assuming, following Feynman, that the integral is exhausted by a single excitation at
$\omega (k)$, one finds $\hbar\omega(k) S(k) = \hbar^2k^2/2m$,
which  yields the dispersion $\omega(k)= \hbar k^2/4m$ in the  molecular regime where $S(k)=2$. 

The possibility of seeing these ``free molecules" in the dynamic structure factor depends in a critical way on the value of 
the dimensionless combination $ka$. In Fig.~\ref{fig4} we show the behavior of the dynamic structure factor as a function 
of $\hbar\omega/\epsilon_F$, where $\epsilon_F=\hbar^2k^2_F/2m$ is the Fermi energy, for a fixed value of momentum transfer 
($k/k_F=3$), so that the energy-weighted integral (\ref{fupup}) of the dynamic structure factor takes the same value for all the 
configurations in the figure. Different values of $k_Fa$ are considered. When $k_Fa$ is positive and small the system is in the 
BEC regime. Fig.~\ref{fig4}a shows that in this case 
($1/k_Fa=2$) most of the strength is concentrated in the molecular peak located essentially at $\omega(k)= \hbar k^2/4m$ 
(corresponding to $\hbar\omega/\epsilon_F= 3^2/2=4.5$). The shift of the position of the peak with respect to the ``free'' 
molecule at
$\hbar k^2/4m$ reflects interaction effects between the molecules. To lowest order it is given by the mean-field result 
$\delta\omega=gn/2\hbar$, where $g=4\pi\hbar^2a_m/2m$ is the molecular coupling constant with $a_m$ denoting the molecule-molecule
scattering length ($a_m=2a$ in the BCS approach, but $a_m=0.6a$ in exact treatments~\cite{Petrov}). For $1/k_Fa=1$ the 
strength in the continuum of atomic excitations is larger, but the molecular peak is still quite strong since it carries 
89.5\% of the total strength. At unitarity (Fig.~\ref{fig4}c) the peak has just merged into the continuum of the single-particle 
excitations, whose spectrum turns out to be strongly affected by molecular-like correlations. Finally if one considers the BCS 
regime of negative 
$k_Fa$ (Fig.~\ref{fig4}d) one finds that rapidly the response is pratically indistinguishable from the one of the ideal Fermi gas 
and no residual effect of interactions is visible at such values of momentum transfer. 

In the same figure we have also shown the prediction for the magnetic dynamic structure factor 
$S_M(k,\omega)= 2[S_{\uparrow\uparrow}(k,\omega) - S_{\uparrow\downarrow}(k,\omega)]$. Compared to $S(k,\omega)$ this 
quantity is not affected by the  presence of the collective mode. Furthermore its strength in the continuum turns out to be 
pushed up to higher frequencies. This is consistent with the fact that $S_M(k,\omega)$ obeys the same f-sum rule (\ref{fupup}) 
as the total structure factor $S(k,\omega)$.

The calculations presented in this work hold for a uniform gas. At the high momentum transfers considered here, the effects 
of trapping, relevant for actual experiments, are expected to give rise to a small Doppler broadening in the molecular peak on 
the BEC side, similarly to what happens in Bose-Einstein condensates~\cite{Bragg}. 
As far as the static structure factor is concerned, the main effects due to the non uniform confinement can be taken into 
account using a local density approximation.  

As already anticipated the dynamic structure factor can be measured through inelastic photon scattering. In order to have 
access separately to the total and spin components of the structure factor various strategies can be pursued. A first 
possibility is to adjust the detuning of the laser light in such a way as to produce a different coupling with the two spin 
components, resulting in a response of the system that will also be sensitive to the spin structure factor. Another possibility
is to work with polarized laser light which can  result in a different coupling in the two atomic spin components even if the 
detuning is large~\cite{Carusotto}.
 
In conclusion we have shown that the structure factor of an interacting Fermi gas at relatively high momentum transfer 
exhibits interesting features associated with the molecular degrees of freedom of the system. In particular in the BEC regime 
and for values of $k$ in the range $k_F<k<1/a$, the total static structure factor approaches the value $2$, characteristic of a 
molecular regime, while the dynamic structure factor is characterized by the occurrence of a sharp peak at 
$\omega(k)= \hbar k^2/4m$. We have found that the up-down spin component of the structure factor exhibits a specific dependence 
on the momentum transfer, being strongly sensitive to the correlations associated with the formation of molecules. Bragg 
scattering experiments to explore the structure factors close to resonance would be highly valuable.

Acknowledgements: SG and SS acknowledge 
support by the Ministero dell'Istruzione, dell'Universit\`a e della Ricerca (MIUR).


\begin{thebibliography}{99}
\bibitem{notemolecule} For a zero-range potential the molecular wave function is given by 
$\Phi_0(r)=\exp(-r/a)/(\sqrt{2\pi a}r)$. This yields the result $S_{\uparrow\downarrow}(k) = (2/ka) \arctan (ka/2)$ which
is very close to the red dotted line in Fig.~\ref{fig1}.
\end{thebibliography}
\end{document}